# Using social annotation and web log to enhance search engine

VU THANH NGUYEN

University of Information Technology
Ho Chi Minh City, VietNam

**Abstract**
Search services have been developed rapidly in social Internet. It can help web users easily to find their documents. So that, finding a best method search is always an imagine. This paper would like introduce hybrid method of LPageRank algorithm and Social Sim Rank algorithm. LPageRank is the method using link structure to rank priority of page. It doesn't care content of page and content of query. Therefore, we want to use benefit of social annotations to create the latent semantic association between queries and annotations. This model, we use algorithm SocialPageRank and LPageRank to enhance accuracy of search system. To experiment and evaluate the proposed of the new model, we have used this model for Music Machine Website with their web logs.

## 1. Introduction

In the period of Internet, search engine is the popular tool and the necessary tool for the web users. However, they have not satisfied when using them such as: difficult to choose the right result in the huge results; difficult to have a right query; difficult to know which results are similar, so on… Therefore, over the past decade, there are many researching to improve the quality of web search. Most of them try to improve some aspects:

1) Rearrange the web pages according to the query document similarity. In this area, some techniques are anchor text generation, metadata extraction, link analysis, search log mining, profile query extraction, profile web usage,..

2) Ordering the web pages according to their priority. It doesn't care the query of user when ranking the priority of web pages. Some techniques are PageRank, HITS,…

In this paper, we optimize web search using LPageRank model by applying social annotations with two aspects: similar ranking and static ranking.

Similar ranking is used to estimate similarity between a query and a web page by annotations. They provide good information to summary of the corresponding web pages. They are metadata which can be used to calculate the similarity between a query and a web page. However, in some web pages, the annotations may be sparse and incomplete. These thing make gap between the annotations and queries. In this paper, we apply an algorithm SocialSimRank (SSR) to enhance this problem.

Another, static ranking is the algorithm ranking priority of web page by structure link.

It uses relationship of web pages in their site to valuate importance of pages. This method must browse all web page to built structure site.

There are many approaches to get the structure site. One of them is to built the structure site with traffic logs which use the information from the website's logs.

Recently, traffic log is used to enhance web search more and more such as: Yahoo patent, LPageRank model of Brin and Motwani and Winograd, LPageRank of Qing Cui and Alex Dekhtyar. However, they often focus one of approach static ranking or similar ranking.

In this paper, the model LPageRank with web log introduced by Qing Cui and Alex Dekhtyar in 2005 and the social annotation information are applied to our model. We have proposed a procedure that computes a score for a web page according to the number of visits to the page and the traffic pattern on the site. We have this hybrid method to build a local search engine. The search engine has been deployed with the web log of website http://machines.hyperreal.org/. The rest of the paper is organized as follows. In Section 2 we describe LPageRank algorithm. In Section 3 we describe social annotation search. In Section 4 we describe the search engine which we have built. In Section 4 we describe our initial experiments and provide the results.

## 2. LPageRank: Using logs in local search

### 2.1 PageRank algorithm

The PageRank algorithm introduced by Brin, Motwani and Winograd is a mechanism in determining the overall importance of a web page. Intuitively, PageRank of a web page is an approximation of a probability to reach this page from some other pages on the web. This computation assigns to each page a PageRank based on the current structure of the website. In the absence of information about human traffic patterns on the web, the PageRank computation assumes that on each page, the user is as likely to follow a specific link, as any other link.

Each link, PageRank assumes that the user is not biased in his/her choice of the link. Thus, the probability to follow a specific link is $\frac{1-\alpha}{m}$. When considering globally, the α leads to the following recursive formula for computing a PageRank (PR) of a page:

$$PR(A) = \alpha + (1-\alpha)\left(\sum_{B \in Parents(A)} \frac{PR(B)}{N(B)}\right) \quad [6]$$

where, Parents(A) is the set of all web pages which link to A and N(B) is the number of outgoing links to distinct pages found on page B. Typically, PageRank is computed iteratively, starting with a state in which it is uniformly distributed among all web pages, and continuing until a stabilization condition holds.

### 2.2 Probabilistic Graph of Web Page Collection

The PageRank implicitly models are used the behavior in terms of a probabilistic graph of the web. Indeed, each page in the collection can be thought of as a node in a graph. A link from page B to page A can be modeled as a directed edge (B, A). Finally, the PageRank computation assumes that, given a page B, the probability to follow some outgoing edge (B, A) is $\frac{1-\alpha}{N(A)}$. The triple G=(W, E, P) where N is a set of nodes, $E \subset WxW$ is a set of directed edges and P:E→[0,1], s.t.,

$(\forall B \in N)\sum_{(B,A) \in E} P(B,A) \leq 1$ is called a *probabilistic graph* [3].

The probabilistic graph constructed (implicitly) by the PageRank computation assumes a uniform probability distribution for outgoing edges for each node.

### 2.3 LPageRank

This method is introduced by Qing Cui and Alex Dekhtyar in 2005. Basically, the LPageRank algorithm is a PageRank computation based on a probabilistic graph of a web page collection that reflects traffic patterns obtained from the logs.

Suppose G= (W, E, P) is a probabilistic graph over a collection of web pages W. Then, LPageRank (LPR) of a web page is computed as follows:

$$LPR(A) = \alpha + (1-\alpha) \sum_{B \in Parents(A)} LPR(B)P(B,A) \quad [3]$$

Note that LPR(A) = PR(A) for graphs G, in which $P(B,A) = \frac{1-\alpha}{N(B)}$ for all edges (B, A).

This model is similar to the Yahoo patent in 2002 as using traffic logs to enhance search engine. However, compared with LPageRank, it has many differences. This method uses web logs to build probabilistic graph and then improves the PageRank algorithm. The URL frequency is not only used to score each URL but also strengthen other URLs which it links to. Moreover, in this model, traffic logs are used to build the structure of website, which is easily than using crawler.

## 3. Search engine with social annotation
### 3.1. Web page annotators

Web page annotators are web users who use annotation to organize memories and share their favorite online. They provide cleaner data which are usually good summarizations of web pages for user's browsing. Besides, similar or closely related annotations are usually given to the same web pages. Base on this observation, SocialSimRank (SSR) is used to measure the similarity between the query and annotations.

In 2007, Shenghua Bao, Xiaoyuan Wu, Ben Fei, Guirong Xue, Zong Su, Yong Yu had an identify about SSR. They consider that SSR is a measure the popularity of web pages from web page annotator's point of view. In figure 1 the Social annotations indicate a, d and c. The a web page is more population than d and c. Each annotation has a relationship with others. The similar (semantically-related) annotations are usually assigned to similar (semantically-related) web pages by users with common interests. In the social annotation environment, the similarity among annotations in various forms can further be identified by the common web pages they annotated.

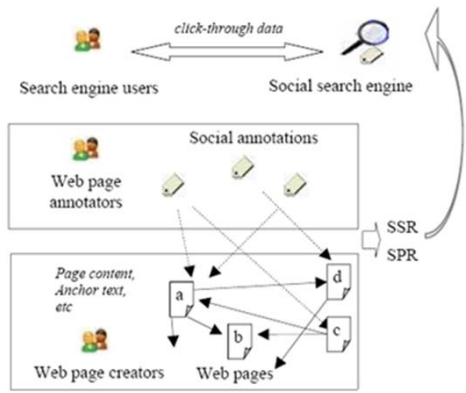

Figure 1 Social annotations

To calculate the similar annotation, we build a relation graph between social annotations and web pages with its edges indicating the user count. Assume that there are *NA* annotations, *NP* web pages and *NU* web users. *MAP* is the $NA \times NP$ association matrix between annotations and pages. *MAP(ax,py)* denotes the number of users who assign annotation *ax* to page *py*. Letting *SA* be the $NA \times NA$ matrix whose element *SA(ai, aj)* indicates the similarity score between annotations *ai* and *aj* and *SP* be the $NP \times NP$ matrix each of whose element stores the similarity between two web pages, we propose SocialSimRank(SSR), an iterative algorithm to quantitatively evaluate the similarity between any two annotations.

Step 1: Init:

Let $S_A^0(a_i, a_j) = 1$ for each $a_i = a_j$ otherwise $0$

$S_P^0(p_i, p_j) = 1$ for each $p_i = p_j$ otherwise $0$

Step 2: Do {

    For each annotation pair $(a_i, a_j)$ do

$$S_A^{k+1}(a_i, a_j) = \frac{C_A}{|P(a_i)||P(a_j)|}$$
$$\sum_{m=1}^{|P(a_i)|} \sum_{n=1}^{|P(a_j)|} \frac{\min(M_{AP}(a_i, p_m), M_{AP}(a_j, p_n))}{\max(M_{AP}(a_i, p_m), M_{AP}(a_j, p_n))} S_P^k(P_m(a_i), P_n(a_j))$$

    For each page pair $(p_i, p_j)$ do

$$S_P^{k+1}(p_i, p_j) = \frac{C_P}{|A(p_i)||A(p_j)|}$$
$$\sum_{m=1}^{|A(p_i)|} \sum_{n=1}^{|A(p_j)|} \frac{\min(M_{AP}(a_m, p_i), M_{AP}(a_n, p_j))}{\max(M_{AP}(a_m, p_i), M_{AP}(a_n, p_j))} S_A^{k+1}(A_m(p_i), A_n(p_j))$$

}

Step 3 Output $S_A(a_i, a_j)$

In this algorithm, *CA* and *CP* denote the damping factors of similarity propagation for annotations and web pages, respectively. *P(ai)* is the set of web pages annotated with annotation *ai* and *A(pj)* is the set of annotations given to page *pj*. *Pm(ai)* denotes the *m*th page annotated by *ai* and *Am(pi)* denotes the *m*th annotation assigned to page *pi*.

Note that the similarity propagation rate is adjusted according to the number of users between the annotation and web page.

Letting $q=\{q1,q2,...,qn\}$ be a query which consists of *n* query terms and $A(p)=\{a1,a2,..., am\}$ be the annotation set of web page *p*,

Equation (4) shows the similarity calculation method based on the SocialSimRank.

### 3.2. Dynamic Ranking page with web log and similarity annotation

Given the probabilistic graph in the previous step, the LPageRank computation is performed. In the first iteration, the LPageRank of each page is allocated by the total scores of all forward links. In the next iteration, the LPageRank of all pages are computed as follows:

$$LPR_{i+1}(A) = \alpha + (1-\alpha) \sum_{B \in Parents(A)} LPR_i(B) P(B,A)$$

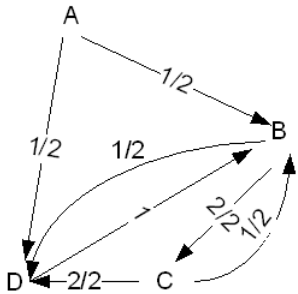

Figure 2 Illustration of quality transition between the users, annotations, and pages in the LPageRank and similarity annotation algorithm

This process is stopped when the difference of LPR$_i$ and LPR$_{i+1}$ is less than small number $\varepsilon$. At the present implementation, $\varepsilon$ is set to 0:0.00001. Then we use the score of each link to calculate the similar annotation.

### 4. Experimental results

Some preliminary experiments have been conducted to test the LPageRank-augmented retrieval. In the following section, the obtained results will be described.

### 5. Experimental Setup

The search engine described in this paper had been implemented on virtual web services and deployed on the local website. All of files of Music Machine Website are downloaded and a virtual website is built such as a online website. To test its performance, we have considered two alternative local search methods: Google's domain-restricted search with URL http://machines.hyperreal.org/ and local PageRank.

Our search engine indexes around 14002 web pages in the http://machines.hyperreal.org/ domain. Only HTML and text files are indexed. For this experiment, the collection of logs is taken from the http://machines.hyperreal.org/ web servers.

Table 1: Summary of result: Google and model local search

| Key words | Google | New model |
|---|---|---|
| Arp-Sequencer | 6 | 2 |
| Roland TR-606 Dramatic | 2 | 2 |
| MonoPoly | 23 | 24 |
| Kawai K3 | 12 | 12 |
| ragtime piano | 1 | 1 |
| Kawai XD5 | 10 | 10 |
| BASS DRUM | 77 | 76 |
| Maplin 3800 | 5 | 6 |
| Maplin 5600S | 7 | 7 |
| Hammond Auto-Vari | 4 | 4 |
| Univox Micro-Rhythmer-12 | 9 | 8 |

A test dataset has been selected consisting of 11 simple keyword queries shown in Table 1.

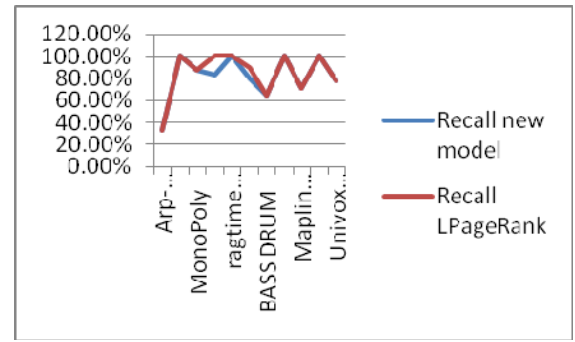

Figure 3: The averaged recall on all query from LPageRank and LPageRank with neural network

## 6. Results

The conclusion is as follows:

- Overall, Google's domain-restricted search is showed the best performance on the test dataset in both precision and average expected precision.
- Local search site has a new order of pages on which the URLs have a high hit weight that often has high rank.
- Some pages disappear from the result of the local search when the hit weight is too low.
- The result of the local search is sensitive with a hit weight.

## 7. Discussion

From the above experiment, two positive conclusions can be reached. First, if everything is given equally, the LPageRank outperforms regular PageRank and provides more relevant links and better quality lists of links to the users. Second, the local search's retrieval appears to be more significantly different than the retrieval by the domain-restricted Google's.

On the other hand, it is a notice that the LPageRank combined by TF-IDF is not enough better than Google's domain-restricted search. There are two potential reasons for Google's better performance: First, while the exact matching formulas of Google are proprietary secrets and ever since we know that Google combines PageRank with more than just TF-IDF based on retrieval. Google's engine analyzes the position of keywords on in the HTML, the text of links pointing to the page, and possibly many other factors ignored by TF-IDF. Second, we have observed that Google has indexed potentially a much larger number of pages in the "http://machines.hyperreal.org/" domain. While some documents indexed by Google are not HTML or text files, and while some other documents are no longer present on the http://machines.hyperreal.org/ web server, it is reasonable to assume that Google's list of http://machines.hyperreal.org/ pages is larger. In part, this can be explained by our crawling mechanism - we will never find any web pages that are NOT reachable from the top page of the website, http://machines.hyperreal.org/. At the same time, Google's crawler is global and can detect these pages if they are linked from outside sources. These results are encouraging and showing that the use of the LPageRank with appropriate retrieval methods may improve retrieval results. They also suggest two avenues for further improvement. First, we must subject our search engine to more rigorous testing, involving independent users rather than authors. Second, we must work to improve the IR component of the search engine and make it resemble more the actual methods used in existing web search engines. We also note that the local search can be used as a means of analyzing the website traffic and changes in it over time.

## 8. Conclusions

In this article, the use of web logs has been studied to argue single website search engines – a type of search engine almost ubiquitously present on all major commerce, academic and interest based on websites. Our preliminary results indicate that our proposed method, local search, outperforms the standard PageRank when coupled with a simple retrieval method (TF-IDF) and clusters of the neural network. An another discovery is that although our local search-based on a search engine shows worse performance than the domain restricted Google search. It helps the users to retrieve significantly different results, and thus shows complimentary to Google behavior. A plan for further studies of the LPageRank and similar annotation has been briefly outlined as well.

In conclusion, the web log resource is used to improve the local search engine, especially the PageRank algorithm and cluster sessions. This model has the high recall but it is still a low process and not a good precision. Although the result is not good, the benefits of web logs in local search engine are discovered, for instance, lower costs in building the structure graph and high reflections of web usages in the result.

**Author Biographies – Vu Thanh Nguyen**

The author born in 1969 in Da Nang, VietNam. He graduated University of Odessa (USSR), in 1992, specialized in Information Technology. He postgraduated on doctoral thesis in 1996 at the Academy of Science of Russia, specialized in IT. Now he is the Dean of Software Engineering of University of Information Technology, VietNam National University HoChiMinh City.

Research: Knowledge Engineering, Information Systems and software Engineering.